# One possible explanation for earthquake occurrences from anomalous line-of-sight propagations in the very high frequency band by fast Fourier transform spectral analysis

Zhao Wang[1], Tie Zhou[2], Kuniyuki Motojima[3], Ze Jin Yang[4]*

*[1] Physical laboratory of Materials Science and Chemical Engineering College, Hainan University, Haikou 570228, China*

*[2] College of Information Science and Technology, Hainan University, Haikou 570228, China*

*[3] Department of Electronic Engineering, Gunma University, 1-5-1 Tenjin-cho, Kiryu 376-8515, Gunma, Japan*

*[4] School of Science, Zhejiang University of Technology, Hangzhou, 310023, China*

**Abstract**

This paper illustrated the possible relationship between the occurrences of the earthquake and the anomalous line-of-sight propagations in the very high frequency band by the fast Fourier transform spectral analysis. Despite many anomalous propagations appear in the different very high frequency band during the earthquake occurrences, the majority of these abnormal signals contain similar frequency distributions in the frequency domain. For the 31 anomalous propagation spectral distributions, 30 of them present the same curve peaks, within a frequency range of $(0\sim0.5)\times10^{-3}$ Hz. Furthermore, for the first time, we found that the spectral maximum of all anomalous propagations are below the characteristic Brunt-Vaisala frequency (period T>6 min), which happens to be the frequency range of the internal gravity waves, which might evidence that the atmospheric gravity waves should be responsible for the indirect coupling between lithosphere and ionosphere. These novel results might provide direct evidence to the relationship between the anomalous propagations in the very high frequency band and the occurrences of earthquakes.

* Corresponding author, Email: zejinyang@zjut.edu.cn

## 1 Introduction

As we all know, it is very difficult to accurately predict the occurrence of Earthquake. In recent years, our understanding on 'Earthquake preparatory activities' have been greatly advanced[1] with the collection of good-quality data. The achieved progress distributed mainly in the non-mechanical field. For example, diverse thermal anomalies including outgoing long-wave radiation, surface latent heat flux, air temperature, relative humidity, and air pressure, and so on, these phenomena usually occurred before seismicity[2-3]. Schekotov *et al*[4] have observed a lithosphere–atmosphere coupling between earthquake and atmospheric tide. The anomalous variance in total electron content of ionospheric during seismicity have been reported in many literatures[5-11]. Pulinets *et al*[12] concluded that the increased radon emanation from active faults and cracks before earthquakes in seismically active areas is the primary source of air ionization. More anomalies on earthquake preparatory activities have been discovered in the field of electromagnetism. For example, the ultra-low frequency (ULF) seismogenic electromagnetic emissions have been observed in some literatures[13-16]. Low frequency (LF) radio signals anomalies associated with earthquakes also have been measured[17-18]. Amplitude and phase anomalies of LF signals have been examined by Rozhnoia *et al*[19] in order to define a threshold of LF signal sensitivity to explain influence of the geomagnetic, solar, and seismic factors. Fukumoto *et al*[20] presented the preliminary results on the possible reception of over-horizon very high frequency (VHF) radio signals from a FM (frequency modulation) transmitter during abnormal situations (probably is closely related with the earthquakes). Anomalous VHF over-horizon signals might relate to impending earthquakes[21-23]. The anomalous propagation in VHF band was observed during earthquakes [24]. Devi *et al*[25] reviewed the relationships between the characteristics of anomalous VHF, received from FM radio transmissions and broadcast television (TV) signals, and the earthquake precursors. These literatures showed that ULF, extremely low, very low/low, medium, high, very high frequency radio waves could be used for short-term earthquake precursor. Therefore, electromagnetic signal indeed can be used to predict the earthquake. To understand earthquake precursor signatures from the lithosphere to the upper ionosphere, the lithosphere–Atmosphere–Ionosphere Coupling model has been discussed in some papers[26-29]. Obviously, substantial progress has been made in non-seismic (electromagnetic) measurements for earthquake precursor during the last decade. Due to the characteristic feature of precursory occurrence and long-distance propagation, Hayakawa and Hobara[30] thought that electromagnetic efforts are decisive superior to conventional seismic measurement. However, the explanations on electromagnetic anomalies induced by earthquake preparation still exist many controversies. The detailed spectrum-related investigations are still absent, which might provide valuable clues according to its profiles during the earthquake occurrences.

The broadcasting waves from Tokyo tower have been monitored continuously since 2007[24]. Some anomalous propagations during the occurrence of earthquakes are observed, which should reflect some information about the earthquakes. This paper focused on the anomalous frequency characteristic and its component analysis by the fast Fourier transform (FFT). These results might deepen our understanding on the relationship between the anomalous signal propagations in the VHF band and the occurrences of earthquakes.

## 2 The brief description of previously completed work

Our measurement systems[24], located at city Kiryu in central Japan, 92 km northwest of Tokyo, have been used to capture the strength of waves propagated on VHF TV band. The target transmitting TV stations locate at Tokyo Tower in Japan. Relative location of transmitter (Tokyo) and receiver at Kiryu is shown in Fig. 1.

The strength of waves in VHF TV band are captured by the measurement system, which consists of multiple antennas, an antenna selector, a spectrum analyzer, a personal computer for data storage, and a web server for open data. All data can be recorded every two minutes. The schematic diagram of the wide band measurement is described in Fig. 2. The target waves, monitored continuously by measurement system, are seven wave values, as are listed in Table 1.

The criterion to distinguish the anomalous data from the normal one is the standard deviation ($\sigma$) of the mean value. Once the signal strength of propagated wave deviation by 3σ or more, the data are regarded as an anomalous propagation. Usually, the specific earthquakes have the magnitude greater than M3.0 and their epicenters were located within about 75km from the propagation path[31]. We select 31 anomalous propagations, all of them are closely related to the occurrence of earthquakes, to deeply analyze their intrinsic relationships. These anomalous propagations are listed in Table 2.

## 3 The fast Fourier transform (FFT) of anomalous propagations

The FFT is an efficient algorithm for computing the discrete Fourier transform (DFT) of a sequence and is also particularly useful in many areas such as signal processing. For a given length *N*, corresponding to a sequence variable *x*, then the DFT could obtain a vector *X* with the same length *N*. FFT and inverse (IFFT) implement the following relationships,

$$X(k) = \sum_{n=1}^{N} x(n) e^{-j2\pi(k-1)(\frac{n-1}{N})} \qquad 1 \le k \le N \tag{1}$$

$$x(n) = \sum_{n=1}^{N} X(k) e^{j2\pi(k-1)(\frac{n-1}{N})} \qquad 1 \le k \le N \tag{2}$$

Important information about a transformed sequence includes its magnitude and phase, which can be calculated by the MATLAB code, in which the magnitude plot is perfectly symmetrical about the Nyquist frequency and the useful information can also be found in the frequency range of 0 to Nyquist.

For our data, the sample frequency is 1/120Hz. The Nyquist frequency is $4.16 \times 10^{-3}$ Hz. The frequency of the FFT can be determined by the following formulas.

$$F_n = (n-1)\frac{F_s}{N} \qquad 1 \le k \le N \tag{3}$$

$$F_s = \frac{1}{120} \tag{4}$$

In order to eliminate the influence of zero frequency, all data all data are obtained by subtracting the average value. Since the appearance of aliasing is possible, we use low-pass filter before FFT. For comparison purposes, 3-hour sampling data have been used to analyze whether it is normal or abnormal. Therefore, the sample length is also long enough to reduce the picket fence effect in FFT. We carried out FFT for 31 anomalous propagations correlated with the occurrence of earthquakes, as is listed in Table 2, with the deviation criterion >3σ. The same analyses were performed for 31 normal propagations until there were no earthquakes in the sensitivity zone of the wave path.

The typical spectral curve of normal propagations is similar to the distribution of the white noise, as is displayed in Fig. 3, that is, there are many peaks and their intensities are comparable, and their positions show more delocalized features and distribute in a wide frequency range. If we define the ratio of the second and the first peak intensities of the curve to characterize the spectral distribution, the ratios are 68% and 96%, respectively. Of the 31 normal propagations, 24 of them are greater than 80%.

Fig. 4 shows the obtained frequency distribution by FFT, for simplicity, only one representative profile is presented, clearly, only one sharp and strong peak is clearly seen, its intensity is far larger than the other ones, displaying a rapid amplitude attenuation with the frequency increasing. In fact, for the 31 anomalous propagations, 30 of them show the peaks within $(0 \sim 0.5) \times 10^{-3}$ Hz, with only one exception of $(1 \sim 2.5) \times 10^{-3}$ Hz. All of the ratios of the second and the first peak intensities are less than 50%. Moreover, if we define

Brunt-Vaisala frequency $\omega_B$ ( $2.78\times10^{-3}$ Hz) as cut-off value to classify the frequency values, the maximum values of all spectra are less than $\omega_B$. The relationship between phase and frequency can also be obtained by FFT for the 31 anomalous propagations. Unfortunately, these relationships failed to provide any new information.

## 5 Discussion

The frequency of VHF band observed in our experiment is smaller than the maximum unstable frequency (MUF) 30MHz, which might not be illustrated by the ionosphere reflection. The major influence on the line-of-sight propagation origins from the changes of the refractive index (n) in the troposphere.

Devi *et al*[1] reviewed the characteristics of anomalous VHF with the earthquake precursors and concluded that the ionospheric effects are less likely in this frequency range due to the limit of MUF. The appearances of anomalous propagation origin mainly from the pre-seismic effects in the troposphere. The previous experimental results evidence the correlations between the lithosphere, the near-surface environment and the VHF propagation characteristics prior to earthquake occurrences. There are several explanations for the coupling mechanism. In fact, to explain all the observations, it is necessary to determine the indirect relationships related to lithosphere-ionosphere coupling than to the electromagnetic or acoustic wave propagation. Molchanov *et al*[27] think only the atmospheric gravity waves (AGW) could be responsible for the indirect coupling. The reduced atmospheric perturbation induced by the temperature and density could follow preseismic water/gas release, further resulting to the generation of internal gravity waves with periods 6–60min, with a frequency range of $2.78\times10^{-3}\sim2.78\times10^{-4}$ Hz . Rozhnoi *et al*[32] found the evident increase in spectral range 10–25min, agreeing with the theoretical estimations on lithosphere-ionosphere coupling by the AGW in the range $\omega<\omega_B$ (period T>6min), where $\omega_B$ is the Brunt-Vaisala frequency.

Our spectrum analysis demonstrates that the frequency distributions of anomalous propagations are indeed different from the normal ones. The former exhibited one strong peak in the low frequency zone, which are uncomparable to the other peaks distributed at relative high frequency zone, whereas the latter contains multiple comparable peaks in a wide frequency range, similar to the distribution of the white noise. Moreover, with the only one exception of the strong peak, all of the peak intensities, including both the anomalous and normal propagations, show generally comparable values. The results indicate that main frequency components distribute in the range of $(0\sim0.5)\times10^{-3}$ Hz in 31 anomalous propagation. The anomalous line-of-sight propagation has less common features in the time domain but more common features in the abnormal frequency domain. More importantly, the maximum of spectrum values are still smaller than $\omega_B$ (period T>6min), where $\omega_B$ is the characteristic Brunt-Vaisala frequency, which happens to be the frequency range of the internal gravity waves. The similar frequency distribution means that most of the anomalous line-of-sight propagation may comply with a similar physical mechanism. These results imply that AGW should be responsible for the indirect coupling between lithosphere and troposphere or ionosphere. The existence of AGW will cause the uneven distribution of the troposphere, which will lead to the change of the refractive index, and ultimately causing the anomalous line-of-sight propagation in VHF band.

Earthquakes preparation process will cause near-surface temperature and density/velocity variations, which are the source of AGW energy. The characteristic frequency of AGW is Brant-Vaisala frequency ($\tau_B$~ 6min) $\omega_B$, which is defined as:

$$\omega_B = \sqrt{\frac{g}{\theta}\frac{\partial\theta}{\partial z}} \qquad (5)$$

Where $g$ is gravity acceleration, $\theta$ is temperature and $z$ is height. There is space-frequency difference of the AGW energy in the atmosphere. The propagation direction of AGW frequency with horizontal plane is defined as angle $\beta$

$$\omega = \omega_B \sin\beta = \sqrt{\frac{g}{\theta}\frac{\partial\theta}{\partial z}} \sin\beta \qquad (6)$$

Our observed data for the earthquake preparatory zones, which is also the source of AGW, are all within a certain distance far away from the propagation path, below about 40 Km for the earthquake at least grade four[31]. For the anomalous propagations, the value of $\sin\beta$ is relatively small, and the observed frequency peak is far less than $\omega_B$. This may also explain why the disturbance would not appear when the epicenter is too far from the propagation path.

## 6 Conclusions

The broadcasting waves from Tokyo tower have been monitored continuously since 2007[24]. The observed data show some anomalous propagations when the earthquake occurrences. In order to find out the valuable information in the anomalous line-of-sight propagation on VHF band, we carried out the frequency spectral analysis for anomalous and normal propagation signals by the fast Fourier transform (FFT) and obtained the following conclusions.

1. The frequency distributions of anomalous propagations are indeed different from the normal propagations. The former exhibited one strong sharp peak at low frequency components, whose intensity is far larger than the other ones, and the latter showed generally comparable peak intensities each other in the whole frequency range, which is also comparable with the ones in the most anomalous cases.
2. Our results show that although anomalous propagations correlate with occurrences of earthquakes in the different carrier wave of the VHF band, whereas the majority of these anomalous propagations contain similar frequency distributions in the frequency domain. Among 31 anomalous propagations frequency distributions, 30 of them present their curve peaks within $(0\sim0.5)\times10^{-3}$ Hz.
3. The maximum values of the most signals spectra are in the range $\omega<\omega_B$ (period T>6min), where $\omega_B$ is the Brunt-Vaisala frequency, which happens to be the frequency range of the internal gravity waves, evidencing that AGW should be responsible for the indirect coupling between lithosphere and troposphere as well as ionosphere. AGW will cause the uneven distribution of the troposphere, which will lead to change of the refractive index, and ultimately causing the anomalous line-of-sight propagation to VHF band.

Our results agree well with the lithosphere- atmosphere-ionosphere coupling by the AGW, and these novel illustrations might provide deep understandings for the earth precursors.

**Acknowledgements**

The authors acknowledge the other members of professor Motojima’s Laboratory for useful discussions on anomalies of seismic activity. Projects supported by the key research and development program of Hainan Province (Grant No. ZDYF2017098) and the natural science foundation of Zhejiang Province (Grant No. LY18E010007).

**Supporting information (SI)**

The supporting information includes five filefolds, with respective names are: (1). 24-hour time domain curve, (2) anomalous propagation spectrum, (3) measurement date, (4) quite date, which means undisturbed data, (5) quite spectrum, which means undisturbed data. The SI also includes a table named as earthquake and anomalous line-of-sight propagation. These original data are for reference purpose.

Table 1. Observed Waves

| Station | NHK General | NHK Education | Nippon TV | TBS | Fuji TV | TV Asahi | TV Tokyo |
|---|---|---|---|---|---|---|---|
| Frequency In MHz | 91.25 | 103.25 | 171.25 | 183.25 | 193.25 | 205.25 | 217.25 |

Table 2. Time of anomaly propagation.

| Date | Anomaly start time(LT) | σ | TV-channel frequency(MHz) | Duration time(minutes) |
|---|---|---|---|---|
| 25/Mar/2007 | 16:00:47 | 4.301 | 205.25 | 59 |
| 10/Apr/2007 | 23:20:09 | -6.178 | 217.25 | 52 |
| 28/Apr/2007 | 3:47:31 | -4.598 | 217.25 | 48 |
| 8/May/2007 | 5:29:32 | -4.487 | 193.25 | 37 |
| 18/May/2007 | 3:17:44 | 4.314 | 103.25 | 45 |
| 19/May/2007 | 16:32:39 | 3.652 | 103.25 | 44 |
| 23/Jun/2007 | 19:24:46 | 4.924 | 103.25 | 179 |
| 16/Aug/2007 | 21:32:47 | -6.916 | 183.25 | 112 |
| 18/Nov/2007 | 8:57:05 | 3.711 | 183.25 | 44 |
| 1/May/2008 | 5:35:02 | -5.759 | 217.25 | 84 |
| 28/May/2008 | 5:51:08 | -3.561 | 103.25 | 33 |
| 12/Jun/2008 | 23:25:55 | -3.945 | 171.25 | 43 |
| 13/Jun/2008 | 2:27:22 | -5.420 | 103.25 | 88 |
| 13/Jun/2008 | 21:39:00 | 3.707 | 217.25 | 118 |
| 15/Aug/2008 | 22:48:21 | -4.683 | 183.25 | 60 |
| 16/Aug/2008 | 0:14:26 | -5.534 | 183.25 | 114 |
| 25/Sept/2008 | 5:37:36 | -4.161 | 193.25 | 53 |
| 26/Oct/2008 | 9:47:08 | 4.684 | 183.25 | 208 |
| 26/Oct/2008 | 13:36:25 | 4.136 | 205.25 | 76 |
| 3/Nov/2008 | 14:37:06 | 3.418 | 217.25 | 45 |
| 3/Nov/2008 | 18:01:10 | 4.613 | 103.25 | 111 |
| 7/Nov/2008 | 10:28:48 | 4.210 | 183.25 | 46 |
| 11/Nov/2008 | 21:25:39 | -6.798 | 193.25 | 124 |
| 20/Mar/2009 | 2:20:05 | -5.617 | 103.25 | 31 |
| 11/Apr/2009 | 5:58:14 | 4.067 | 183.25 | 152 |
| 13/May/2009 | 3:26:49 | -6.036 | 217.25 | 156 |
| 16/Aug/2009 | 19:07:02 | 4.586 | 91.25 | 234 |
| 7/Nov/2009 | 1:33:53 | -4.130 | 183.25 | 56 |
| 8/Nov/2009 | 23:29:38 | -4.830 | 183.25 | 53 |
| 28/Jan/2010 | 16:41:28 | -5.926 | 183.25 | 36 |
| 10/Mar/2010 | 2:24:42 | -4.829 | 217.25 | 33 |

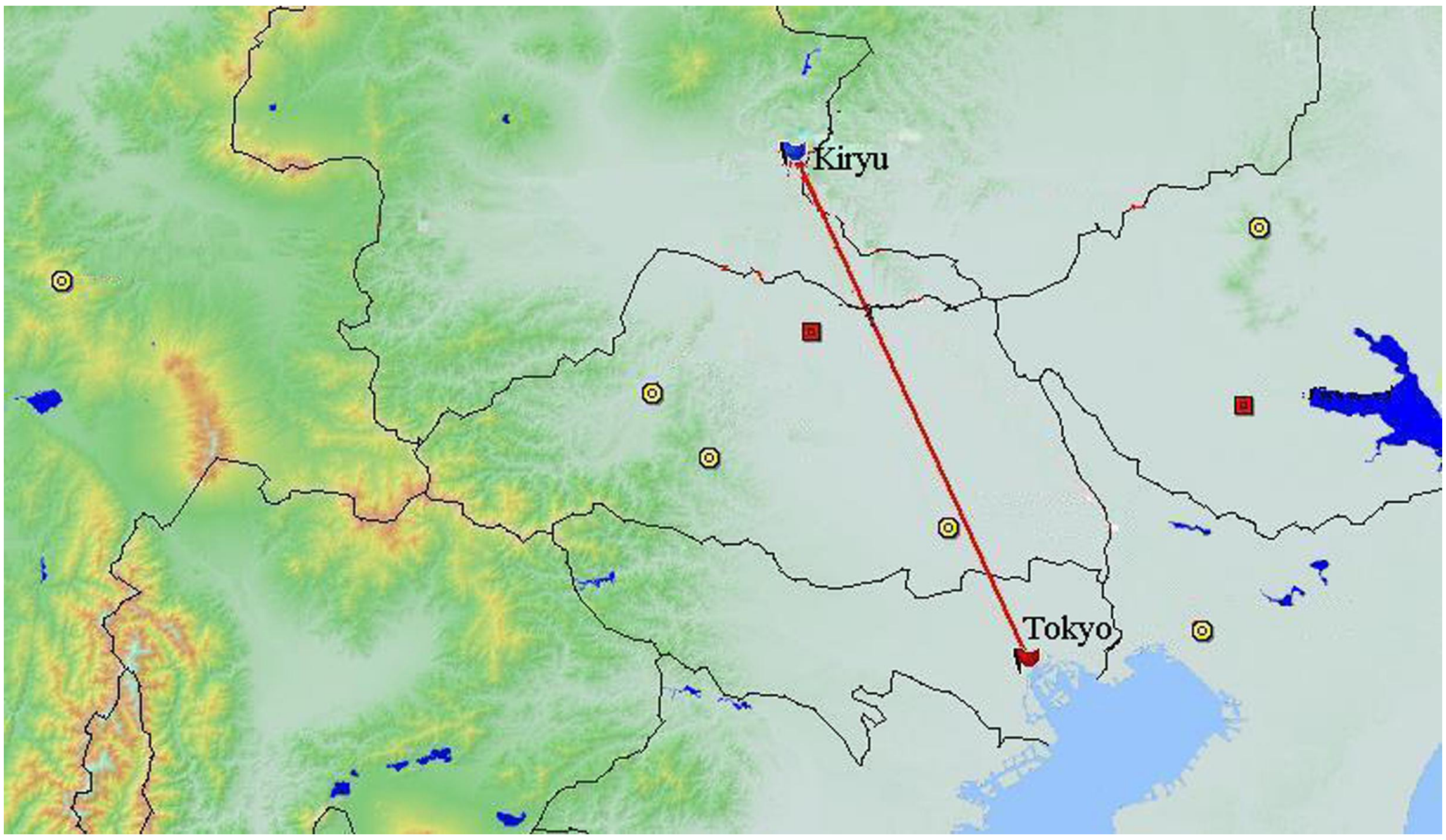

Fig.1. Relative location of transmitter (Tokyo) and receiver (Kiryu).

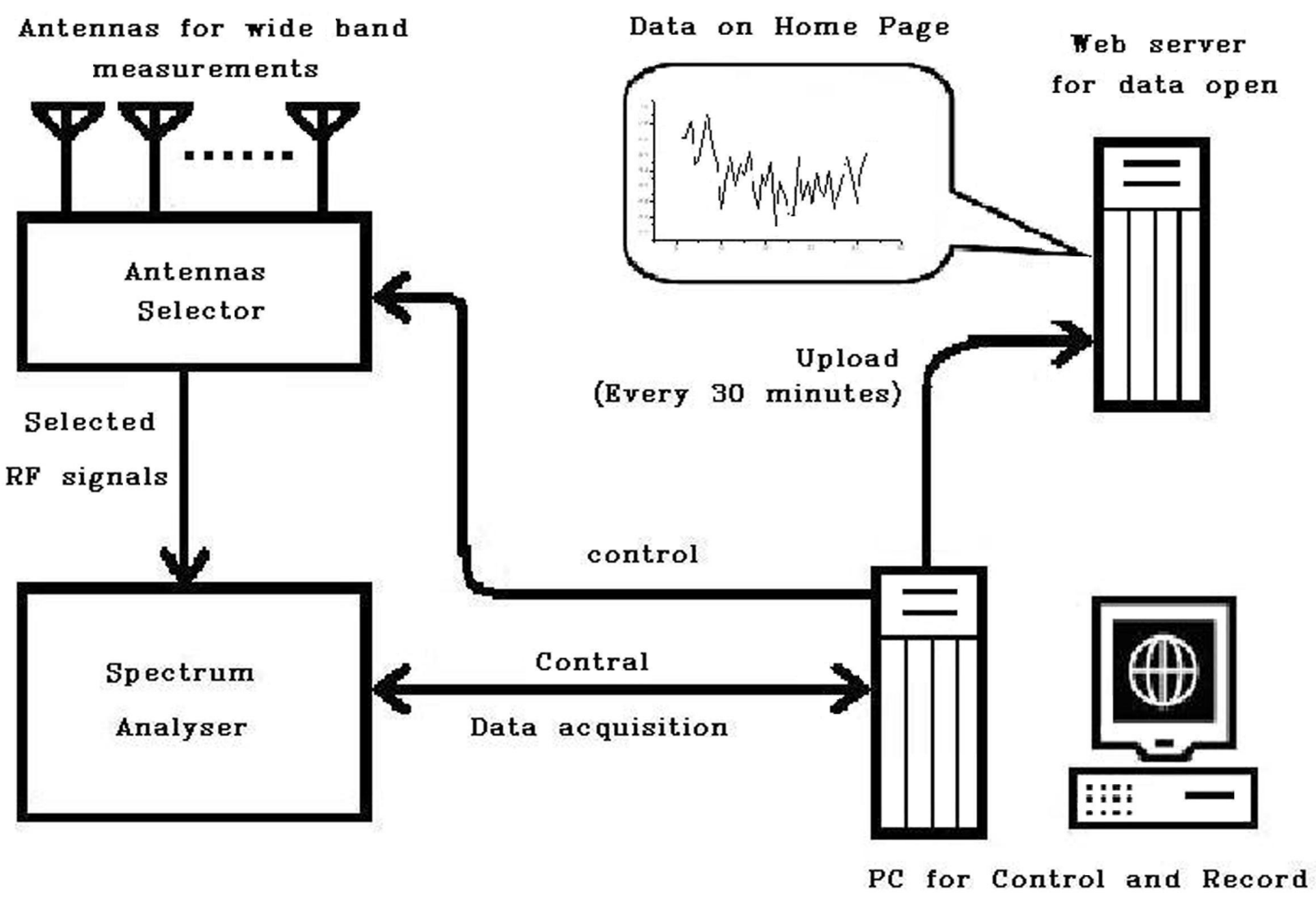

Fig.2. Schematic diagram of the wide band measurement system.

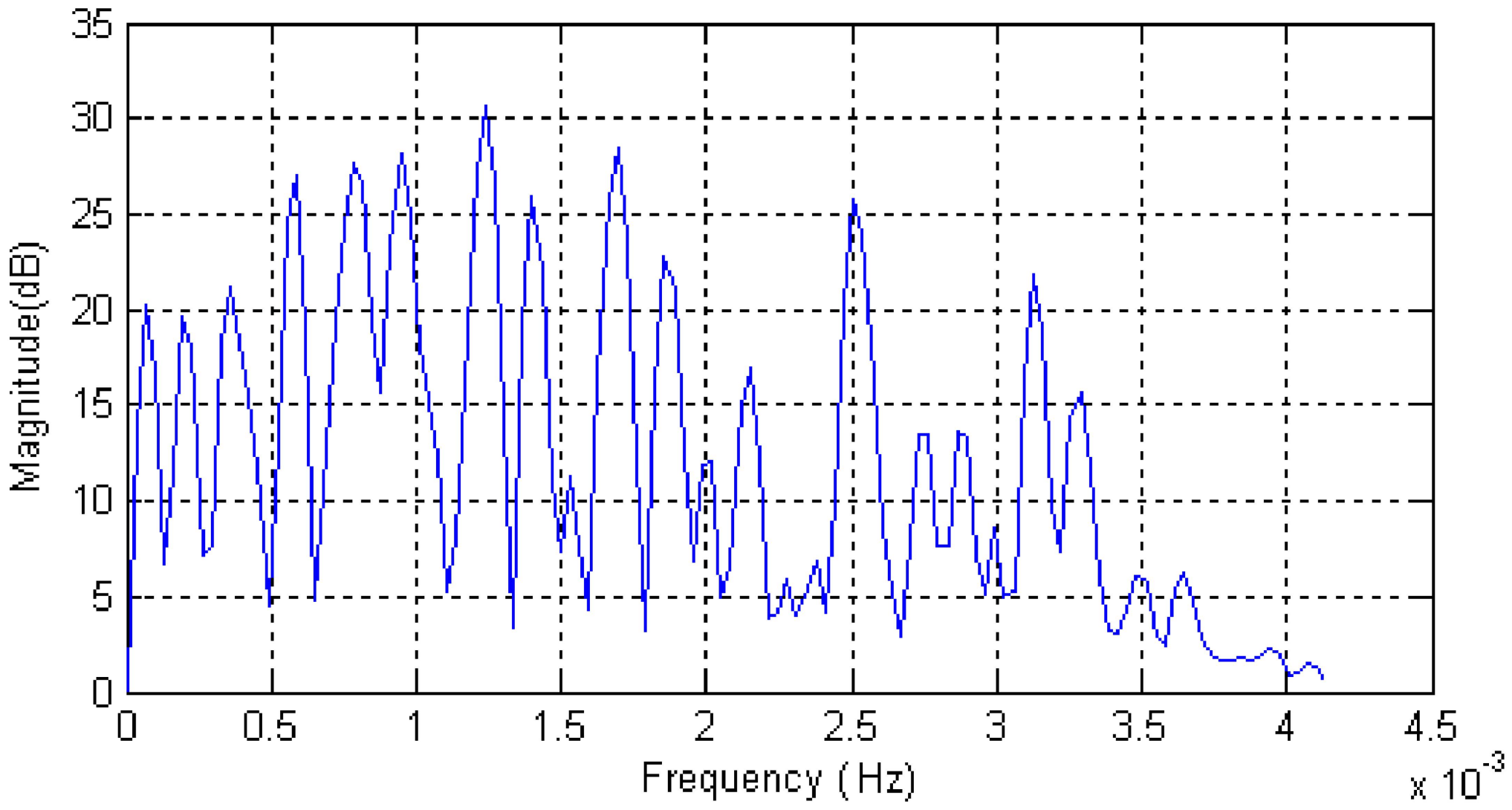

Fig. 3. Frequency distribution of normal propagations signal.

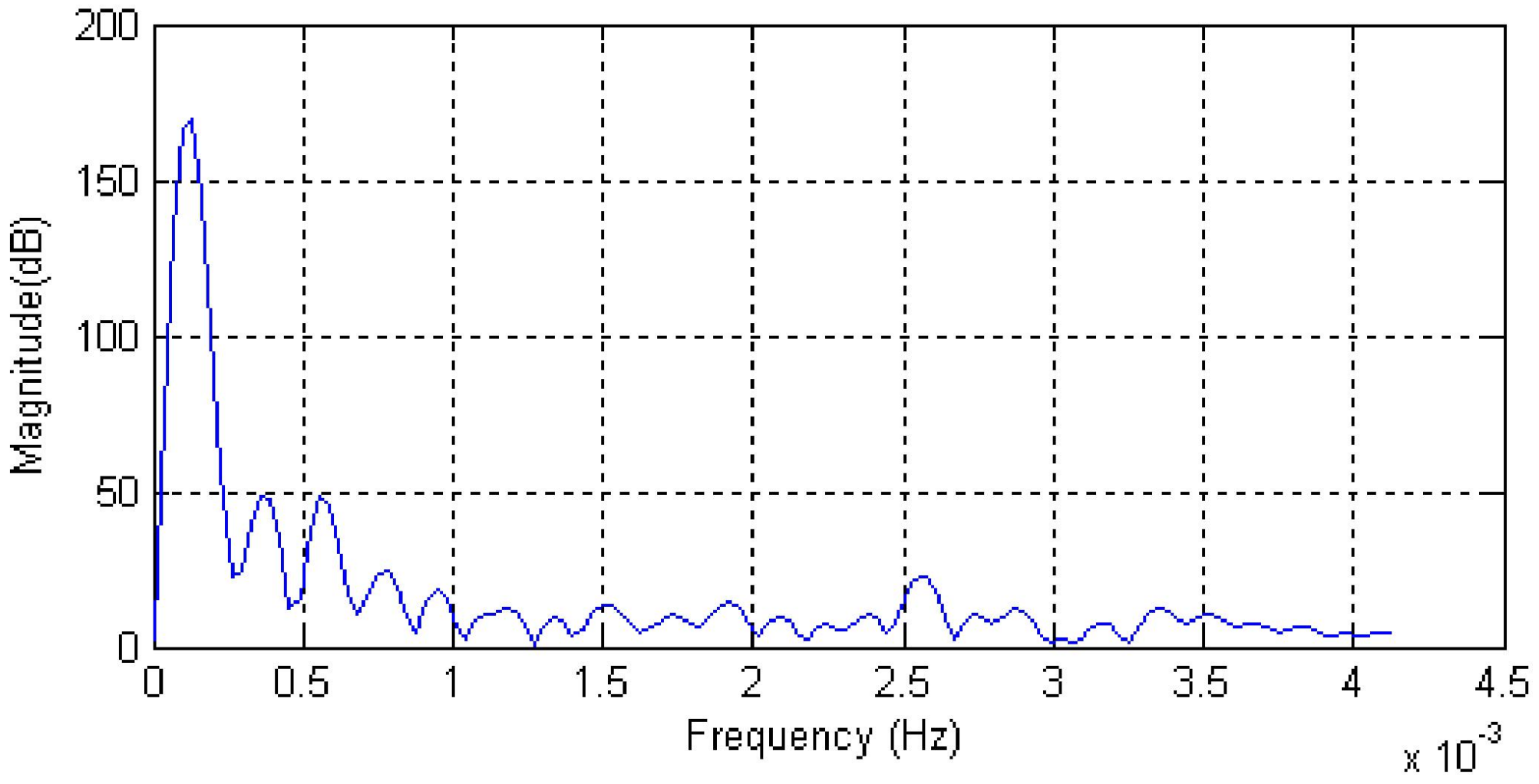

Fig.4. Anomalous propagations frequency distribution.